# TTC: A Tensor Transposition Compiler for Multiple Architectures

Paul Springer
AICES, RWTH Aachen University, Germany
springer@aices.rwth-aachen.de

Aravind Sankaran
RWTH Aachen University, Germany
a.sankaran@grs-sim.de

Paolo Bientinesi
AICES, RWTH Aachen University, Germany
pauldj@aices.rwth-aachen.de

## Abstract

We consider the problem of transposing tensors of arbitrary dimension and describe TTC, an open source domain-specific parallel compiler. TTC generates optimized parallel C++/CUDA C code that achieves a significant fraction of the system's peak memory bandwidth. TTC exhibits high performance across multiple architectures, including modern AVX-based systems (e.g., Intel Haswell, AMD Steamroller), Intel's Knights Corner as well as different CUDA-based GPUs such as NVIDIA's Kepler and Maxwell architectures. We report speedups of TTC over a meaningful baseline implementation generated by external C++ compilers; the results suggest that a domain-specific compiler can outperform its general purpose counterpart significantly: For instance, comparing with Intel's latest C++ compiler on the Haswell and Knights Corner architecture, TTC yields speedups of up to $8\times$ and $32\times$, respectively. We also showcase TTC's support for multiple leading dimensions, making it a suitable candidate for the generation of performance-critical packing functions that are at the core of the ubiquitous BLAS 3 routines.

*Categories and Subject Descriptors* D.1.3 [*Concurrent Programming*]: Parallel programming; G.4 [*Mathematical Software*]: Parallel and vector implementations; I.1 [*Symbolic and Algebraic Manipulation*]: Languages and Systems—Special-purpose algebraic systems

## 1. Introduction

Tensor transpositions are an important building block for tensor contractions,[1] which appear in a wide range of applications such as machine learning [1, 17], quantum chemistry calculations [2, 6], multidimensional Fourier transforms [4, 13] and climate simulations [3].

---

[1] From a computational perspective, tensors can be viewed as multidimensional arrays, and tensor contractions can be thought of as a generalization of a matrix product.

To make efficient use of the highly tuned *Basic Linear Algebra Subprograms* (BLAS) libraries, tensor contractions can be cast in terms of general matrix-matrix multiplications via tensor transpositions. However, the existence of a high-performance tensor transposition kernel is critical to render this approach useful. Due to the non-contiguous memory access patterns and the vast number of architecture-specific optimizations required by modern vector processors (e.g., vectorization, blocking for caches, non-uniform memory accesses (NUMA)), writing high-performance tensor transpositions is a challenging task. Until now, many research efforts focused on 2D [5, 9, 11, 12, 18] and 3D transpositions [8, 15], while higher dimensional transpositions [10, 19] are mostly still uncovered.

To this end, we created the *Tensor Transposition Compiler* (TTC), a domain-specific compiler which automates the task of generating high-performance, vectorized and parallelized code for multiple architectures. To make it applicable to a wide range of applications, we designed TTC to support transpositions and additions of the form

$$B_{\Pi(i_0,i_1,...,i_{N-1})} \leftarrow \alpha \times A_{i_0,i_1,...,i_{N-1}} + \beta \times B_{\Pi(i_0,i_1,...,i_{N-1})}, \quad (1)$$

where $A$ and $B$ are $N$-dimensional tensors,[2] $\Pi(i_0, i_1, ..., i_{N-1})$ denotes an arbitrary permutation of the indices $i_0, i_1, ..., i_{N-1}$, and $\alpha$ and $\beta$ are scalars. Hence TTC is able to generate transpositions which overwrite (or update) the output tensor $B$ with a transposed (and scaled) input tensor $A$. If $\alpha = 1$ and $\beta = 0$, then Eqn. 1 reduces to an ordinary out-of-place tensor transposition.

In comparison to our previous publication on TTC [14], this work focuses on the usage of TTC as a tool; for instance, we outline its application to the generation of packing routines encountered in high-performance BLAS implementations. Moreover, we now support CUDA-enabled GPUs as well as Intel's Xeon Phi coprocessors; and while Intel's next generation Xeon Phi, Knights Landing (KNL), is not yet released, we also support AVX512-enabled processors.[3] Finally, we assess TTC's performance with two different compilers, and stress the importance of proper thread affinity.

The remainder of this paper is structured as follows. Section 2 summarizes the code-generation process of TTC and gives exemplary use cases. Section 3 evaluates the performance of TTC on five different architectures (i.e., Intel Haswell, Intel Xeon Phi, AMD Steamroller, NVIDIA Kepler, NVIDIA Maxwell), shows speedups over a relevant baseline and illustrates the importance of thread affinity. Section 4 draws conclusions.

---

[2] We adopt the Fortran memory layout, storing the tensor indices from left to right (i.e., column-major).

[3] The correctness of the code for KNL was verified via Intel's software development emulator.

## 2. Tensor Transpose Compiler

TTC[4] is an open-source compiler, written in Python, which generates high-performance C++/CUDA C code for arbitrarily dimensional tensor transpositions. For any given transposition and tensor size, TTC explores a search space of different implementations, henceforth called "candidates". These candidates differ in their characteristics (such as loop order or blocking), which we briefly discuss in Section 2.2. A detailed description of the generation process of TTC is beyond the scope of this paper; the interested reader is referred to [14] for further information. In this paper we concentrate on the capability and features of the tool.

### 2.1 Overview

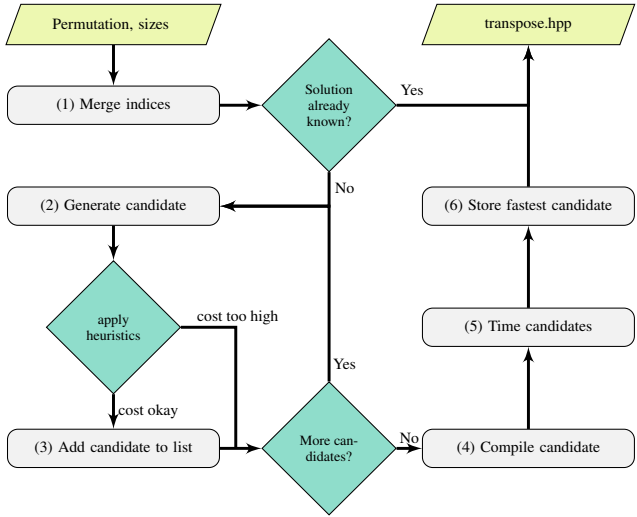

**Figure 1:** Schematic overview of TTC; vectorization and parallelization are always enabled.

Fig. 1 summarizes the code-generation process. As input, TTC expects a symbolic representation of the tensor transposition as well as the tensor size (see top left of Fig. 1). To reduce complexity, TTC merges neighboring indices whenever possible (stage 1). For instance, given the permutation $\Pi(i_0, i_1, i_2) = (i_1, i_2, i_0)$, the indices $i_1$ and $i_2$ are merged into a new 'super index' $\tilde{i_1} := (i_1, i_2)$ of the same size as the combined indices (i.e., size($\tilde{i_1}$) = size($i_1$)×size($i_2$)); as a consequence, the permutation becomes $\Pi(i_0, \tilde{i_1}) = (\tilde{i_1}, i_0)$.[5] Next, TTC queries a local SQL database of known/previous implementations, to check whether an implementation for the input transposition and size already exists; if so, no generation takes place, and the previous implementation is returned. Otherwise, the generation process starts.

Throughout the generation process, TTC maintains a list of candidates. Users can set the capacity of this list via the command-line argument --maxImplementations (see Table. 1); by default, the capacity is set to 200 candidates. In stage (2), TTC explores all candidates (i.e., all combinations of loop orders and blockings); a candidate is added to the list (stage 3) only if the list contains fewer than maxImplementations candidates, or if the estimated cost of the current candidate (according to an internal heuristic) is lower than the highest estimated cost of the worst candidate in the list. In the latter case, the worst candidate (highest estimate) is replaced by the current candidate.

---
[4] Code is available at www.github.com/HPAC/TTC

[5] Two indices $i_m$ and $i_{m+1}$ can be merged if and only if ld($i_m$) = size($i_m$), where size($i$) and ld($i$) respectively denote the size and leading-dimension of index $i$.

Once all candidates have been ranked, TTC continues to (4) compile and (5) time the most promising candidates (i.e., those candidates from its list). Finally, in stage (6) the fastest candidate (henceforth called "solution") is selected and stored to a *.hpp* file, while its timing information and its characteristics (e.g., blocking, loop order, see Section 2.2) are saved for future reference in the SQL database.

| Argument | Description |
|---|---|
| --perm=<index1>,<index2>,... | permutation |
| --size=<size1>,<size2>,... | size of each index |
| --dataType=<s,d,c,z,sd,ds,cz,zc> | data type of $A$ and $B$ |
| --beta=<value> | beta |
| --lda=<lda1>,<lda2>,... | leading dimensions of $A$ |
| --ldb=<ldb1>,<ldb2>,... | leading dimensions of $B$ |
| --maxImplementations=<value> | max #implementations |
| --architecture=[avx,cuda,knc] | architecture |
| --compiler=[g++,icpc] | external C++ compiler |
| --numThreads=<value> | number of threads |
| --affinity=<string> | thread affinity |

**Table 1:** A subset of TTC's command-line arguments; the only two required arguments are --perm and --size.

Table 1 lists a subset of TTC's command-line arguments.[6] Given the desired permutation[7] and tensor size, TTC generates C++ code for the target architecture. All common numerical data types are supported (i.e., single, double, single-complex and double-complex); additionally, mixed-precision transpositions are also supported, allowing different data types for the input tensor $A$ and the output tensor $B$. For instance, by specifying the command-line argument --dataType=sd, TTC generates a mixed-precision transpositions for which $A$ is stored in single-precision (denoted by 's') while $B$ is stored in double-precision (denoted by 'd').

By means of the arguments --lda=... and --ldb=..., TTC supports multiple leading dimensions for each index of $A$ and $B$. This feature enables users to operate on sub-tensors as part of a larger tensor (see Fig. 2a). One important application of this feature is the generation of high-performance packing kernels which are frequently used by level 3 BLAS operations such as a matrix-matrix multiplication. Moreover, thanks to TTC's mixed-precision support, those level 3 BLAS routines can be extended to mixed-precision effortlessly [16].

Figure 2 illustrates an exemplary use case of TTC, showing the packing and transposition of a sub-tensor $A_{i_0,i_1,i_2}$ into a contiguous tensor $B_{i_2,i_1,i_0}$. Notice that if the optional arguments --lda=... and --ldb=... are omitted, the respective tensors are assumed to be stored contiguously; thus, in this example $B$ is stored contiguously. The input to TTC for such a packing request is outlined in Fig. 2b. Figure 2c shows the interface of the generated transposition. Despite the fact that the solution (i.e., the fastest candidate) was generated for a specific size (this is reflected in the name of the transposition), it can be safely used for any other size as well; however, since this might yield suboptimal performance, we encourage users to generate a new solution whenever the size of the target tensor is substantially different than that of a previous solution. We opted to provide the size arguments as template parameters as opposed to function arguments to give the C++ compiler more optimization opportunities (e.g., most of the loop trip counts are known at compile time)—if necessary, this design choice can easily be changed.

---
[6] The complete list of arguments can be displayed via ttc --help.

[7] We use the terms permutation and transposition interchangeably.

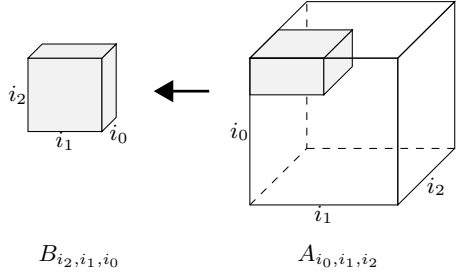

(a) Visualization of permutation.

```
ttc --perm=2,1,0 --size=8,16,16 --lda=32,32,32
```

(b) Input to TTC.

```
/**
 * B(i2,i1,i0) <- alpha * A(i0,i1,i2)
 *
 * \param[in] A Input tensor
 * \param[out] B Output tensor
 * \param[in] alpha scalar factor of A
 * \param[in] lda leading dimensions of A, can be NULL
 * \param[in] ldb leading dimensions of B, can be NULL
 */
template<int size0, int size1, int size2>
void sTranspose210_8x16x16(const float* A, float* B, float
    alpha, const int *lda, const int *ldb);
```

(c) Generated C++ interface.

**Figure 2:** Example. Packing and transposition of a non-contiguous sub-tensor $A$ into a contiguous, packed tensor $B$.

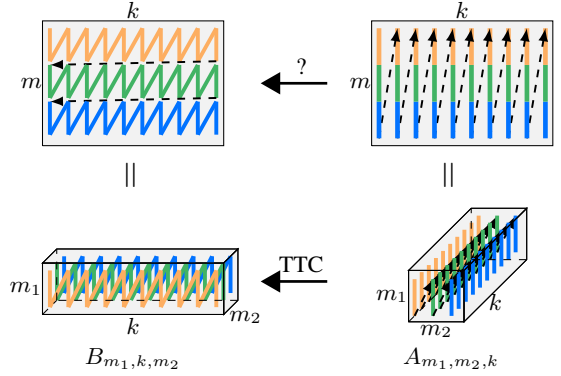

(a) Visualization of a packing operation.

```
ttc --perm=0,2,1 --size=S_{m_1},S_{m_2},S_k
```

(b) Input to TTC.

**Figure 3:** Example. Packing common in BLAS calls. Contiguous lines denote memory layout (arrows into the third dimension are omitted for better readability). $S_{m_1}, S_{m_2}, S_k$ respectively denote the size of the indices $m_1$, $m_2$ and $m_3$.

Figure 3a (top) depicts a typical packing example encountered in BLAS 3 routines [16]. Given a 2D input tensor (top right), in order to best fit the underlying architecture, one has to rearrange the data into a different layout (top left). To achieve such a packing via TTC, one reinterprets the input matrix (2D tensor) $A_{m,k}$ as a 3D tensor $A_{m_1,m_2,k}$ (bottom right), so that the total size of the $m$ dimension does not change (i.e., size($m$) = size($m_1$)size($m_2$)). Likewise, the output matrix $B_{m,k}$ (top left) can reinterpreted as a 3D tensor $B_{m_1,k,m_2}$ (bottom left). It is important to notice that this alternative representation of a 2D tensor as a 3D tensor (or vice versa) solely changes the symbolic representation, but not the memory layout. The input to TTC to generate such a packing request is shown in Fig. 3b.

### 2.2 Vectorization and Parallelization

The key design decision behind TTC is to break an arbitrarily dimensional tensor transposition down into multiple, independent 2D transpositions; Fig. 4 illustrates this process in the 2D case. We refer to these independent transpositions as macro-tiles (see dark green, blue and orange tiles at the bottom of Fig. 4). Those macro-tiles are processed in parallel by different threads (threadblocks) in the context of x86-based (CUDA-based) systems.

The macro-tiles internally consists of several micro-tiles (shown in shades of green for the dark green macro-tile in Fig. 4). Each micro-tile represents a fully-vectorized in-register (in-shared-memory) transposition written in intrinsics (CUDA C) for any x86-based (CUDA-based) architecture. By representing an arbitrary tensor transposition in form of these micro-tiles, we are able to abstract most of the architectural differences away, such that the code generation process remains mostly unchanged between different architectures. The value of $w$ is chosen to match the vector width of the specified architecture and the selected data type (e.g., $w = 8$ for single-precision transpositions running on an AVX-enabled processor). The only processor dependent informations utilized by TTC are the vector width and the size of each cacheline; more precisely, knowledge about the sizes of the caches or the number of TLB entries is not exploited.

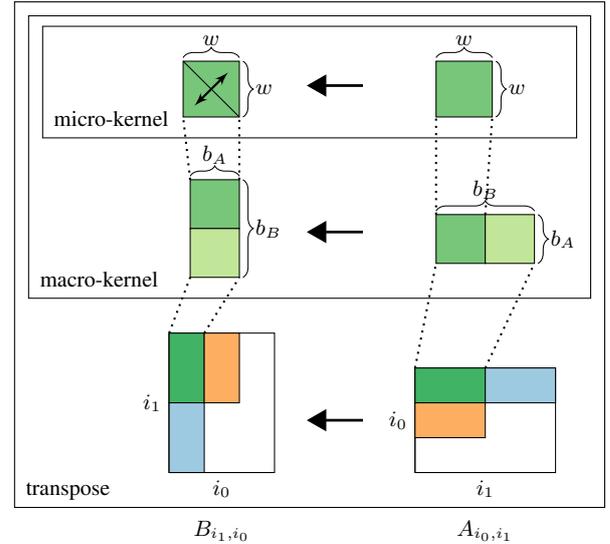

**Figure 4:** Decomposition of a 2D transpose into macro-tiles and micro-tiles.

TTC generates different candidates based on the values of $b_A, b_B \in \{w, 2w, 3w, 4w\}$ as well as the order in which the macro-tiles are scheduled among the threads/threadblocks. For a $d$-dimensional transposition, all $d!$ loop orders are explored to schedule the macro-tiles. For instance, looking at Fig. 4, TTC can choose to schedule first either the macro-tiles along the $i_0$ or those along the $i_1$ dimension.

## 2.3 Reduction of Search Space

Since the number of possible loop orders grows very quickly with the dimension of the tensors, TTC applies heuristics to prune the search space of possible candidates. Aiming to identify a good loop order, "loop-heuristic" favors loop orders with small strides for the accesses to $A$ and $B$ in the innermost loop. This choice increases spatial locality and thus helps to reduce TLB and cache misses; it also increases the likelihood that modern hardware prefetchers prefetch cache lines from main-memory to the caches.[8]

In addition to the loop-ordering, TTC also applies a heuristic to choose suitable values for $b_A$ and $b_B$. On the one hand, to prevent false sharing of cache-lines between threads and to make transfers from caches to main-memory as efficient as possible, this heuristic favours values of $b_A$ and $b_B$ which are multiples of the cache-line size. On the other hand, it aims to minimize the "remainder" of the tensor; the remainder $r^T$ of tensor $T$ is defined as $r^T = S_1^T \pmod{b_T}$, where $S_1^T$ denotes the size of the stride-1 index of $T$. In the current version of TTC, the transposition of $r^T$ is handled via scalar loops; in the future we will explore the possibility of using small(er) block-sizes to vectorize even these leftovers.

## 3. Performance Evaluation

We evaluate the performance of TTC on five different platforms: KNC, HSW, STR, K40 and M840 (see Table 2). Each measurement reports the minimum runtime out of three runs; the caches are cleared before each run. The number of threads for the given systems are chosen such that they yield optimal performance. To establish an upper bound for the performance, Table 2 also presents the *peak*, i.e., the bandwidth attained by the optimized SAXPY.[9] If not otherwise mentioned, we use a compact thread placement: pinning logically neighboring threads to physically neighboring cores, ignoring hyperthreads. All the arrays are initialized in parallel to distribute the data evenly among the memory controllers in a NUMA environment.[10]

To assess the performance of TTC across a wide range of use cases, we report TTC's bandwidth on a synthetic tensor transpositions benchmark [14].[11] The benchmark comprises a total of 57 transpositions ranging from 2D to 6D; each tensor of the benchmark is of size 200 MB.

The bandwidth of a TTC-generated solution $\chi$ is computed as

$$\text{Bandwidth}(\chi) := \frac{3 \times S}{1024^3 \times \text{Time}(\chi)} \text{ GiB}/s, \qquad (2)$$

where $S$ denotes the size of the transposed tensor (in bytes). The prefactor 3 comes from the fact that the output tensor $B$ is updated (i.e., $\beta \neq 0$).

### 3.1 Bandwidth

Figure 5 summarizes the performance of TTC's generated solutions on all five platforms. The number of candidates is limited to 1, 10, 100 or $\infty$ (i.e., no limit; provided via `--maxImplementations=-1`). In all platforms, TTC's solutions—limited to 100 candidates—achieve a significant fraction of peak memory bandwidth; the performance for HSW (Fig. 5b) and M840 (Fig. 5e) are especially impressive, as all the solutions across the benchmark attain 90+% efficiency. To put these results into perspective, a recent study on random tensor permutations by Lyakh [10] presents results between 30 and 55 GB/s, and between 10 and 33 GB/s for an Intel KNC system and an NVIDIA K20X system, respectively.[12] Since it is not known either which exact permutations are considered, or how the measurement is performed, these results should not be understood as a one-to-one comparison; however, they give an idea of the potential of TTC.

Panels 5b (HSW) and 5e (M840) suggest that TTC's heuristics work so well that the search could almost be avoided. The story, however, changes drastically for the KNC platform (see Panel 5c); here, several candidates have to be evaluated before a high-performance implementation is identified. This points to the fact that there might be an architectural feature which is not covered by the heuristics, thus leaving room for future investigations. With the exception of some outliers, the performance on the STR (see Fig. 5a) and K40 (see Fig. 5c) systems is also stable.

### 3.2 Speedup

Figure 6 shows the speedup of TTC's solution over a reference implementation. The reference for a $d$-dimensional transposition consists of $d$ perfectly nested loops annotated with `#pragma omp parallel for collapse(`$d$`-1)` on the outermost loop, as well as `#pragma omp simd` on the innermost loop; moreover, the loops are ordered such that the output tensor $B$ is accessed in a perfectly linear fashion to avoid false sharing of cache lines among different threads.[13] All in all, the compiler was helped as much as possible to generate efficient code.

The speedups over the baseline vary greatly from platform to platform and from test case to test case. As it is evident from Fig. 6c, the KNC platform experiences the greatest speedups, up to $32\times$. While the speedups for the STR (see Fig. 6a) and HSW (Fig. 6b) system are more moderate, they still outperform the external C++ compiler by up to $19.44\times$ and $8.84\times$, respectively.

The benchmark also exposes cases for which TTC does not deliver any appreciable speedup over the baseline. As one can infer from Fig. 5, this is not a shortcoming of TTC, but it is instead an indication that in those cases the baseline is already very good. A closer look at those test cases reveals that they (mostly) correspond to transpositions for which the stride-1 index does not change (e.g., $B_{i_0,i_2,i_1} \leftarrow A_{i_0,i_1,i_2}$, $B_{i_0,i_3,i_2,i_1} \leftarrow A_{i_0,i_1,i_2,i_3}$). The reason for the good performance in these cases is that the external C++ compiler can copy a whole column at a time (instead of single elements), hence avoiding costly scattered memory accesses.

### 3.3 Effect of Thread Affinity and Compilers

Figure 7 compares the bandwidth across the benchmark on the HSW system for two different compilers and two different thread placements/affinities: "compact" and "scatter". The former places logically neighboring threads as close together as possible—without running on the same physical core—while the latter places them as far apart as possible.

The results indicate that the compiler has only a marginal effect on the achieved performance (compare ▲, ▲); this is expected since the innermost kernel is written entirely in AVX intrinsics. The thread affinity, on the other hand, affects performance severely

---

[8] For instance, the maximal stride supported by hardware prefetchers of Intel's Sandy Bridge CPUs is limited to 2 KiB [7].

[9] As defined by the BLAS, SAXPY is the single-precision vector-vector addition ($\mathbf{y} \leftarrow \alpha \mathbf{x} + \mathbf{y}, \alpha \in \mathbb{R}, \mathbf{x}, \mathbf{y} \in \mathbb{R}^n$).

[10] Linux applies the *first touch policy*, meaning that data is allocated close to the thread which touches the data first—not the thread who allocates the data.

[11] The complete benchmark is available at www.github.com/HPAC/TTC/tree/master/benchmark

[12] In our experiments, the K20X has a peak SAXPY bandwidth of 157 GiB/s.

[13] Interestingly, for the KNC architecture this loop order proved to be suboptimal; hence, we chose a loop order which traverses the input tensor $A$—instead of $B$—in a linear fashion.

| Name | Microarchitecture | Model | #Cores | #Threads | ECC | SAXPY [GiB/s] | Compiler | Compiler-flags |
|------|-------------------|-------|--------|----------|-----|---------------|----------|----------------|
| STR  | AMD Steamroller   | A10-7850K | 4 | 2 | Off | 18.05 | g++ 5.3 | -O3 -march=native |
| HSW  | Intel Haswell     | E5-2680 v3 | $2 \times 12$ | $2 \times 12$ | On | 103.98 | icpc 16.0.1 | -O3 -xHost |
| KNC  | Intel Knights Corner | 5110p | 240 | 120 | On | 139.43 | icpc 16.0.1 | -O3 -mmic |
| K40  | NVIDIA Kepler     | Tesla K40c | 2880 | 256* | On | 156.7 | nvcc 7.0.27 | -O3 -arch=sm_35 |
| M840 | NVIDIA Maxwell    | GeForce 840m | 384 | 256* | Off | 12.45 | nvcc 7.5.17 | -O3 -arch=sm_50 |

**Table 2:** Platforms used. The number of threads were chosen to yield the best performance on the given platform.   *: threads per threadblock.

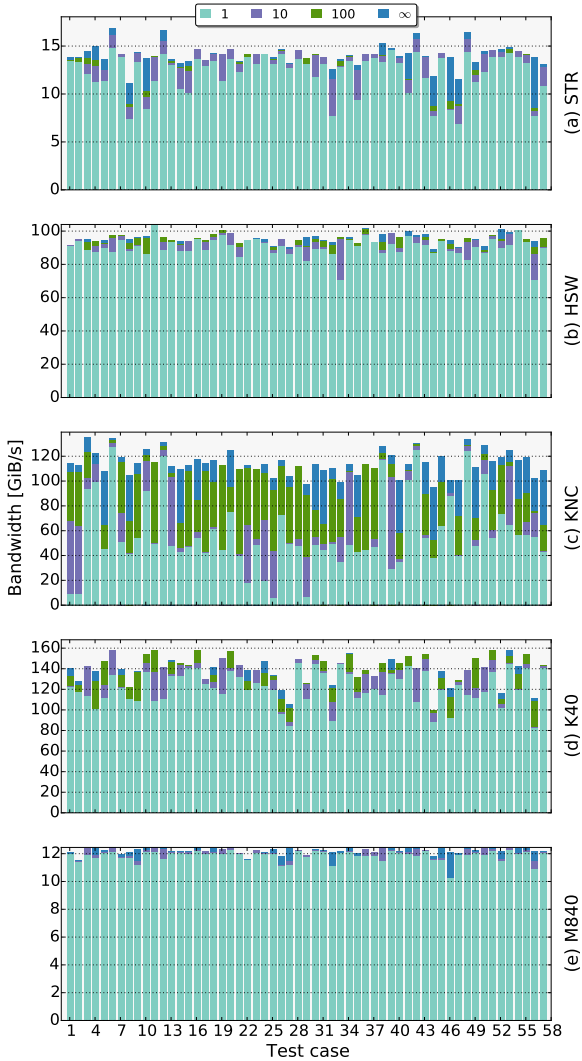

**Figure 5:** Bandwidth for each test case of the benchmark. The top of each plot denotes the SAXPY bandwidth.

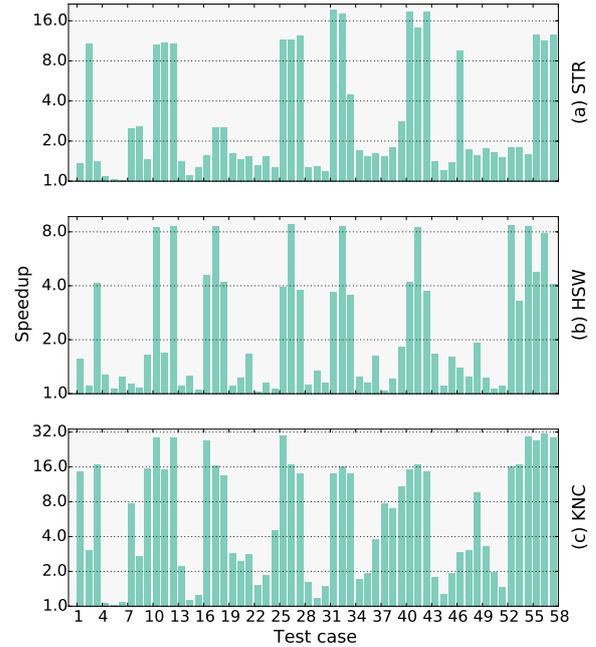

**Figure 6:** Speedup over reference implementation across the benchmark.

(compare △, ○ and ▲, ●), favouring a "compact" thread placement over a "scattered" thread placement by up to $35\%$ (see test case 55).

Given that thread placement has such a noticeable impact on performance, users of TTC are encouraged to specify the desired thread affinity via the `--affinity` command-line argument. Notice that the numbering of cores might change from one system to another.

## 4. Conclusions

We presented TTC, a multi-threaded tensor transposition compiler. On all tested platforms, TTC achieves a substantial fraction of the peak memory bandwidth and significantly outperforms general purpose C++ compilers. We demonstrated the strength of TTC's heuristics to prune the search space of possible candidates, effectively reducing the compilation time. Our evaluation of the effects of thread affinity on performance suggests that in the context of tensor transpositions a compact thread affinity is superior. By supporting multiple leading dimensions, it is possible to operate on sub-tensors; because of this feature, TTC is used as a building block for a tensor contraction compiler that we are currently developing.

### Acknowledgments

Financial support from the Deutsche Forschungsgemeinschaft (DFG) through grant GSC 111 is gratefully acknowledged.

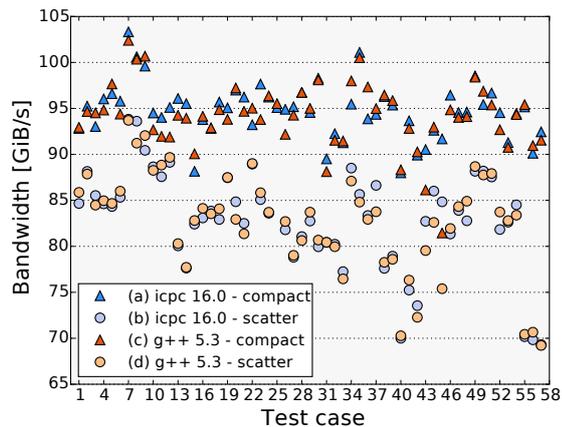

**Figure 7:** Compiler and thread affinity on HSW system.